\pdfoutput=1
\documentclass[aps,prl,reprint,amsmath,amssymb,footinbib,superscriptaddress]{revtex4-1}
\usepackage{custom}

\begin{document}

\title{Evolution of Non-Gaussian Hydrodynamic Fluctuations}
\author{Xin An}
\email{xan@unc.edu}
\affiliation{Department of Physics, University of Illinois, Chicago, Illinois 60607, USA}
\affiliation{Department of Physics and Astronomy, University of North Carolina, Chapel Hill, North Carolina 27599, USA}

\author{G\"{o}k\c{c}e Ba\c{s}ar}
\email{gbasar@unc.edu}
\affiliation{Department of Physics and Astronomy, University of North Carolina, Chapel Hill, North Carolina 27599, USA}

\author{Mikhail Stephanov}
\email{misha@uic.edu}
\affiliation{Department of Physics, University of Illinois, Chicago, Illinois 60607, USA}
\affiliation{Kadanoff Center for Theoretical Physics, University of
Chicago, Chicago, Illinois 60637, USA}

\author{Ho-Ung Yee}
\email{hyee@uic.edu}
\affiliation{Department of Physics, University of Illinois, Chicago, Illinois 60607, USA}

\date{\today}

\newpage

\begin{abstract}   

  In the context of the search for the QCD critical point using
  non-Gaussian fluctuations, we obtain the evolution equations for
  non-Gaussian cumulants to the leading order of the systematic expansion
  in the magnitude of thermal fluctuations.  We develop a diagrammatic
  technique in which the leading order contributions are given by tree
  diagrams.  We introduce a 
  Wigner transform for {\em multipoint} correlators and derive the
  evolution equations for three- and four-point Wigner functions for
  the problem of nonlinear stochastic diffusion with multiplicative
  noise.

\end{abstract}

\maketitle

\section{Introduction}
\label{sec:intro}

The recent resurgence of interest in the classic subject of
hydrodynamics in general~\cite{Landau:2013fluid} and hydrodynamic
fluctuations~\cite{Landau:2013stat2} in particular has been largely
driven by the progress in heavy-ion collision experiments that create
and study droplets of hot and dense matter governed by the physics of
strong interactions described by quantum chromodynamics (QCD).  The
importance of fluctuations in heavy-ion collisions is due to the fact
that the QCD fireballs created in such experiments, with typical
particle multiplicities $\mathcal O(10^{2-4})$, while being large
enough for hydrodynamics to apply, are not too large for fluctuations
to be negligible.

Such fluctuations are observable in heavy-ion collisions via
event-by-event measurements. Furthermore, fluctuations are enhanced if
the matter created in the collisions is in a state close to a critical
point and can serve as signatures of the
criticality~\cite{Stephanov:1998dy,Stephanov:1999zu,Stephanov:2008qz,Stephanov:2011pb}
in the beam energy scan
experiments~\cite{Aggarwal:2010cw,Bzdak:2019pkr}.  The magnitude of
the signatures is determined by the competition between the critical
slowing down and the finiteness of the expansion
time~\cite{Stephanov:1999zu,Berdnikov:1999ph,Mukherjee:2015swa,Akamatsu:2018tcp}.

This necessitates a quantitative description of the fluctuation {\em evolution\/}
within a hydrodynamic framework, and there have been significant
advances in that area recently
\cite{Kapusta:2011gt,Kapusta:2012zb,Young:2014pka,Stephanov:2018hydro+,Akamatsu:2017,Akamatsu:2018,Martinez:2018,Nahrgang:2018afz,An:2019rhf,An:2019fdc,An:2020jjk,Rajagopal:2019hydro,Du:2020bxp,Chao:2020kcf}. Most
relevant for this work is the formalism describing the evolution of
correlation functions coupled to the hydrodynamic background. 
While the approach was considered long ago in a nonrelativistic context~\cite{Andreev:1978}, the relativistic formalism has been introduced recently in the boost-invariant Bjorken flow characteristic
of heavy-ion collisions in Refs.~\cite{Akamatsu:2017,Akamatsu:2018}, and in general
background in Refs.~\cite{An:2019rhf,An:2019fdc}.

However, so far  the formalism  has been limited to {\em
  two}-point correlation functions. The description of the
higher-point correlators quantifying the {\em non-Gaussianity} of the
fluctuations has been elusive until now.
On the other hand, the experimental search for the QCD critical point
relies heavily on such measures of non-Gaussianity
\cite{Stephanov:2008qz,Stephanov:2011pb,Bzdak:2019pkr,Aggarwal:2010cw}
(which, similar to fluctuation magnitudes,  depend on
time evolution \cite{Mukherjee:2015swa}).  We  tackle this
crucial gap
between the ability of the theory w.r.t. non-Gaussian fluctuations
and the demand of the experiment.

We obtain evolution equations for appropriate measures of
non-Gaussianity in the hydrodynamic regime, that is, the regime where the ratio
of correlation length to typical fluctuation wavelength is small. Such
a regime exists even near the critical point, provided the correlation
length remains much shorter than the size of the system, as is the
case in most physical systems, including heavy-ion collisions.

\section{General multivariable formalism}
\label{sec:generic-form}

To understand better the issues of nonlinearity and multiplicative
noise, which is essential for non-Gaussian fluctuations, we begin with a more
general formalism for a {\em discrete} set of stochastic variables
$\sq_i$, labeled by index $i$. The set of stochastic Langevin
equations reads
\begin{equation}
  \label{eq:qdot}
\frac{d \sq_i}{dt} = F_i[\sq] + H_{ij}[\sq]\noise_j\,,
\end{equation}
where drift $F$ and noise magnitude $H$ are functions of 
$\sq_i$, summation over repeated indices is implied, and $\noise_i$ is the
Gaussian white noise, i.e.,
\begin{equation}
  \label{eq:xixi}
  \langle\noise_i(t_1)\noise_j(t_2)\rangle = 2\delta_{ij} \delta(t_1-t_2).
\end{equation}

Eq.~(\ref{eq:qdot}) suffers from a well-known ambiguity
often referred to as the problem of multiplicative
noise: 
it needs
additional information to define the product of the stochastic function
$H[\sq]$ and the noise~$\noise$. From our point of view, this is not
really a problem but rather a shortcoming of the notation used in
Eq.~(\ref{eq:qdot}), which does not reflect the necessary
information.  The ambiguity is removed by discretizing time in
Eq.~(\ref{eq:qdot}) and taking the limit $\Delta t\to0$. The choice of
the definition of the product of $H$ and~$\noise$ is a matter of
convenience. We shall use the choice known as Ito calculus, where $H$
and~$\noise$ are evaluated at the same moment.  It is also well-known
that the change of the definition of stochastic calculus (e.g.,
Stratonovich instead of Ito) is simply equivalent to a shift of the
drift term $F_i$~\cite{ZinnJustin:2002ru}.

Using Ito calculus, one can derive a 
Fokker-Plank equation for the probability distribution of $\sq$:
\begin{equation}\label{FP_gauge_Ito}
  \partial_t P=
\left(-F_i P + \left(Q_{ij}P\right)_{,j}\right)_{,i}
,
\end{equation}
where $Q\equiv HH^T$,
$ (\ldots)_{,i}\equiv\partial(\ldots)/\partial \q_i$ and
$\partial_t\equiv\partial/\partial t$. The Fokker-Plank equation is
unambiguous as written, 
unlike the
Langevin equation~[Eq.~\eqref{eq:qdot}].

Let us consider the equilibrium solution $ P_{\rm eq}$ to the Fokker-Plank
equation, i.e., $\partial_t P_{\rm eq}=0$. While the divergence of the
probability flux on the rhs of Eq.~(\ref{FP_gauge_Ito}) vanishes,
the flux itself does not have to and could be equal to the divergence of
an antisymmetric 2-form, which we write as
\begin{equation}
  \label{eq:F=omega}
  F_i P_{\rm eq} - \left(Q_{ij}P_{\rm eq}\right)_{,j} 
= \left(\Omega_{ij} P_{\rm eq}\right)_{,j}\,,
\end{equation}
where $\Omega_{ij}=-\Omega_{ji}$.
Thus, $F_i$ can be expressed in terms of the equilibrium distribution
\begin{equation}
  \label{eq:F=Peq}
  F_i = P_{\rm eq}^{-1}\left((Q+\Omega)_{ij}P_{\rm eq}\right)_{,j}   
= M_{ij}S_{,j} + M_{ij,j}\,,
\end{equation}
where we introduced the Onsager matrix
\begin{equation}
  \label{eq:M}
  M \equiv Q + \Omega \quad\mbox{and also}\quad S \equiv \log P_{\rm eq}\,.
\end{equation}

Rather than defining the stochastic process in terms of the function
$F_i$, which additionally needs specification of the stochastic
calculus rule, it makes more sense to use $S$ and $M$ to define the
process~\cite{Arnold:1999uza}.
While $S$ describes the equilibrium distribution, and in a certain
sense is analogous to entropy, the matrix~$M$ describes the dynamics
of the stochastic process: the symmetric semipositive definite part
$Q$ is responsible for relaxation and the antisymmetric matrix
$\Omega$ (symplectic form) for the Hamiltonian-like nondissipative motion.
 These {\em physical} properties of the process are independent of the
stochastic calculus prescription, while their relationship to drift~$F$
depends on such a prescription and is written in Eq.~(\ref{eq:F=Peq})
for Ito calculus.

\section{Perturbative expansion}
\label{sec:pert-expans}

Using the Fokker-Plank equation [Eq.~\eqref{FP_gauge_Ito}], we can now write the
evolution equation for any function of variables~$\sq$, including
arbitrary products of their fluctuations $\delta \q_i=\sq_i-\av{\sq_i}$
or $n$-point functions such as
$G_{i_1\ldots i_n}\equiv\av{\delta \q_{i_1}\ldots \delta
  \q_{i_n}}\equiv G_n$.

For example, if $S$ is a
bilinear function of $\q_i$ and~$M_{ij}$ are constants, (i.e., when Eq.~(\ref{eq:qdot}) describes a linear Ornstein-Uhlenbeck process) the equations for $G_n$
are linear and involve only $G_n$ and $G_{n-2}$, and therefore can be solved
iteratively. The equations for the {\em cumulants},
\begin{equation}
  \label{eq:cumulants}
  G^\tc_{i_1\ldots i_n}
  \equiv \left.\frac{\partial^n\log\av{\exp(\mu_i\sq_i)}}{\partial\mu_{i_1}\ldots\partial\mu_{i_n}}\right|_{\mu=0},
\end{equation}
are even simpler --- they decouple from each other.
Equilibrium is achieved when all $n>2$ cumulants vanish, as expected for the
Gaussian distribution $P_{\rm eq}=e^S$, while
$G^{\rm eq}_{2}=-(S'')^{-1}$ where $(S'')_{ij}\equiv S_{,ij}$.

In general,  $S$ is not a bilinear and $M_{ij}$ are not
constants, and we have an infinite system of coupled equations for
cumulants. To organize this system into a hierarchy, we develop a
perturbation theory. 

In many physical systems of interest, 
particularly in hydrodynamics, the fluctuations around
the equilibrium are {\em controllably} small. In other words,
the probability distribution $P$ is sharply peaked and  
can be treated as Gaussian in the lowest order of an approximation.

To be 
systematic, we 
introduce an expansion parameter $\varepsilon$ to
control the magnitude of the deviations
from equilibrium. 
We assume
\begin{equation}
  \label{eq:Seps}
  S'' \sim \varepsilon^{-1}
\end{equation}
and that this remains true for all higher-order derivatives of $S$,
ensuring that for small $\varepsilon$ the probability distribution~$e^S$
approaches a narrow Gaussian with the characteristic magnitude of
fluctuations  $|\delta \q|\sim\sqrt{G_2}\sim\sqrt\varepsilon$.

Thus,
\begin{math}
  G_{2n-1}\sim G_{2n} \sim \varepsilon^{n}.
\end{math}
The odd-order moments $G_{2n-1}$ are of the same order as $G_{2n}$ because $\langle\delta \q\rangle=0$. On the other hand, the non-Gaussian
{\em cumulants} are smaller for the same order:
\begin{equation}
  \label{eq:Gcneps}
  \Gc_{n} \sim \varepsilon^{n-1}.
\end{equation}
This power counting is easily established by a
diagrammatic expansion of Eq.~(\ref{eq:cumulants}) using probability~$e^S$, with $(S'')^{-1}\sim\varepsilon$
playing the role of a propagator and each vertex being of order
$\varepsilon^{-1}$. 

\begin{figure}[ht]
  \centering
  \includegraphics[scale=0.2]{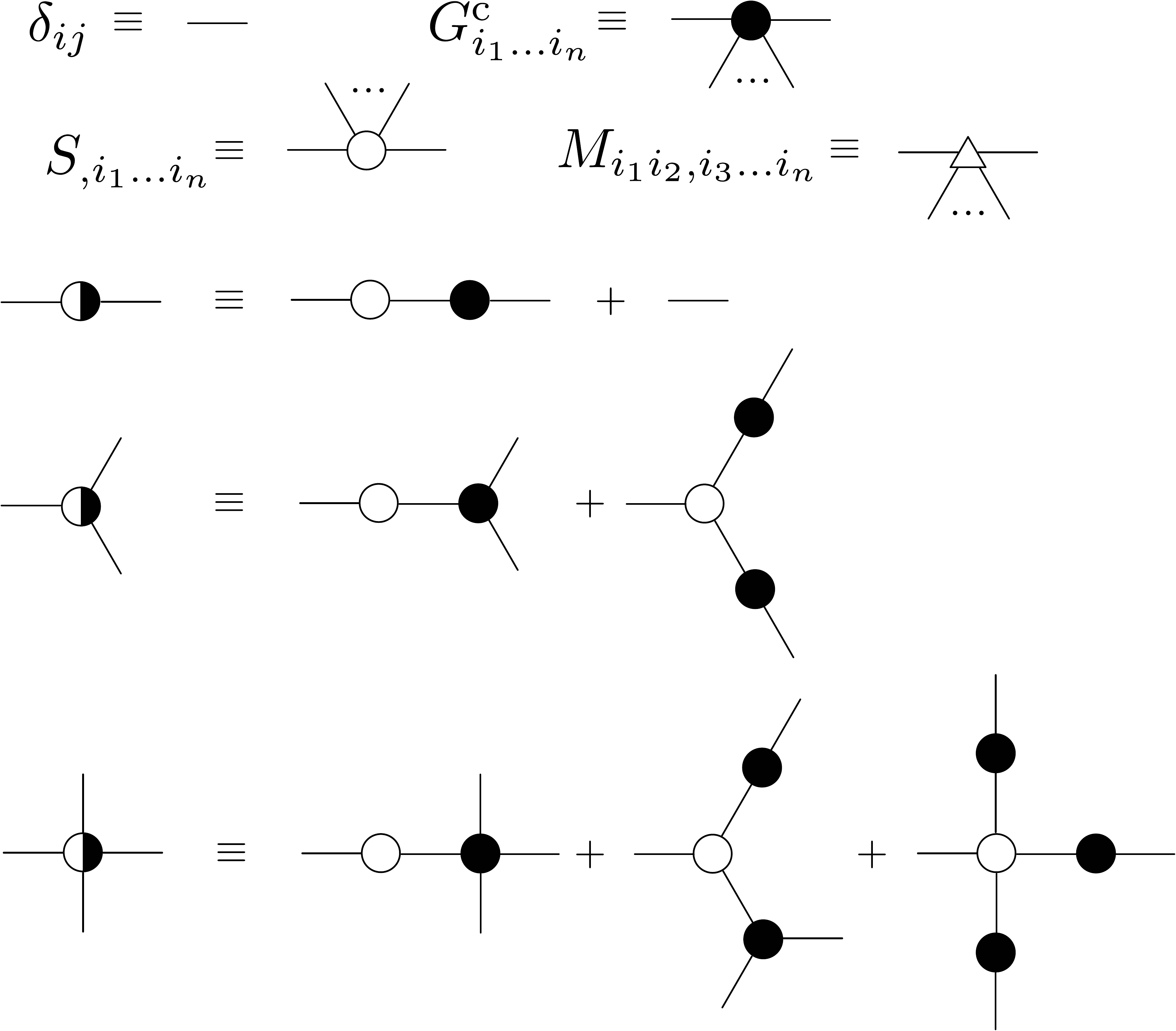}
  \caption{Diagrammatic notation used in
    Fig.~\ref{fig:diagram_equations} to represent the evolution
    equations [Eq.~\eqref{eqs:dt_Wc234}]. The subsets of diagrams
    denoted by half-filled circles vanish in equilibrium.}
  \label{fig:diagram_notations}
\end{figure}

\begin{figure}[ht]
  \centering
  \includegraphics[scale=0.2]{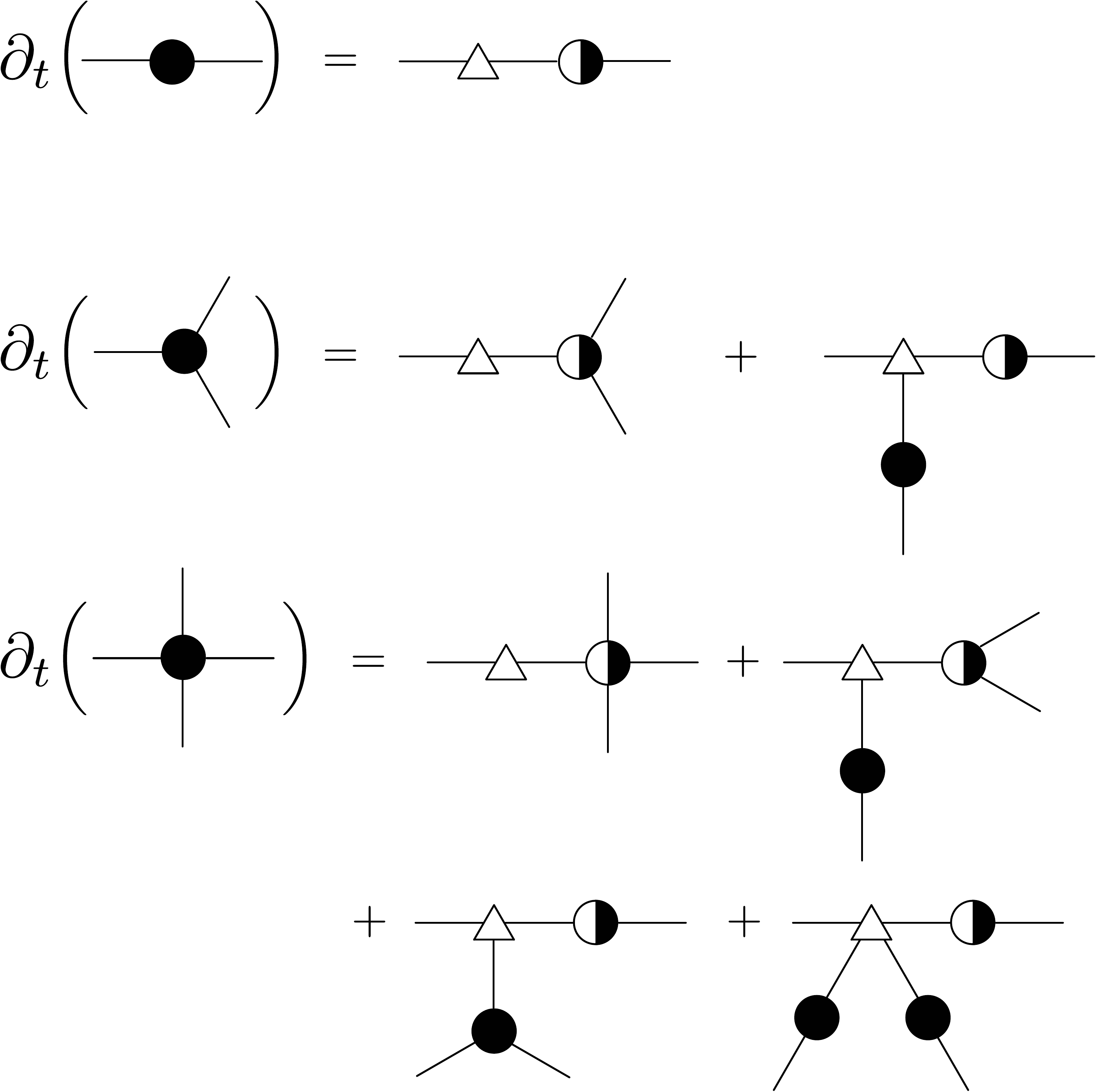}
  \caption{Diagrammatic representation of the evolution equations
    [Eq.~\eqref{eqs:dt_Wc234}] using notations described in Fig.~\ref{fig:diagram_notations}.}
  \label{fig:diagram_equations}
\end{figure}

Because, according to Eq.~(\ref{eq:Gcneps}), cumulants of higher
orders are progressively suppressed, the hierarchy is now
parametrically controllable by $\varepsilon$. Truncating each equation
at the leading order,
we find, for $n=2,3$,

\begin{subequations}
  \label{eqs:dt_Wc234}
\begin{equation}
\del_t \Gc_{i_1i_2}
  = \left[\M_{i_1j}\left(S_{,jk}\Gc_{ki_2}+\delta_{ji_2}\right)\right]_\ipermii
                ,\label{eq:dt_Wc2_mv}
\end{equation}
\begin{multline}
\del_t \Gc_{i_1i_2i_3}
=\Big[\frac{1}{2}\M_{i_1j}\left(S_{,jk}\Gc_{ki_2i_3}+S_{,jk\ell}\Gc_{ki_2}\Gc_{\ell i_3}\right)
\\
+\M_{i_1j,m}\Gc_{mi_2}\left(S_{,jk}\Gc_{ki_3}+\delta_{ji_3}\right)
\Big]_\ipermiii , \label{eq:dt_Wc3_mv}
\end{multline}
\end{subequations}
where $[\ldots]_\ipermn$ denotes the sum over permutations of indices.
The corresponding equation for $\del_t \Gc_{i_1i_2i_3i_4}$ contains nine terms
(before index permutations), and it is easier to express using
diagrammatic representation in Figs.~\ref{fig:diagram_notations}
and~\ref{fig:diagram_equations}. The symmetry factors, such as $1/2$ in
Eq.~(\ref{eq:dt_Wc3_mv}), count the number of permutations that do
not produce different terms.

It is notable that the only diagrams appearing at the leading order in
$\varepsilon$ are ``trees.'' The first term in each equation is the same
as in the linear problem, as it involves
only $S''$ and no derivatives of $M$. Other terms are due to
nonlinearities and/or multiplicative noise, which contribute at the
same order in $\varepsilon$.  Higher-order terms contain loop
diagrams, which we shall not discuss in this Letter.

In most applications, the noise $\noise$ is almost Gaussian,
being a cumulative result of many uncorrelated factors. In our
approach, one can treat noise non-Gaussianity systematically when $H$
is a small parameter and the noise cumulants obey the same hierarchy,
as in Eq.~(\ref{eq:Gcneps}). These assumptions hold in hydrodynamics.
The $n$-th 
($n>2$) noise cumulant contribution will enter at order $H^n$, while the
leading terms we keep in Eq.~(\ref{eqs:dt_Wc234}) are of order $H^2$.




Eq.~(\ref{eqs:dt_Wc234}) and its diagrammatic representation are
among the main results of this work. We shall apply
this approach to hydrodynamics, i.e., a system of stochastic
continuum fields, rather than
discrete variables.

\section{Stochastic Nonlinear Diffusion}

As the simplest example of a stochastic hydrodynamic problem, we shall
consider the diffusion of conserved density (of, e.g., particle number or
charge) in a slowly evolving (e.g., expanding and/or cooling)
medium. This problem carries the most important features we want to
address in this work, including multiplicative noise, without
the complications of additional degrees of freedom and the Hamiltonian
(nondissipative) dynamics of ideal fluid motion. We shall postpone
the extension of these results to full stochastic hydrodynamics to
future work but will discuss the implications of our findings  below.

The diffusion equation is essentially a conservation equation
for fluctuating density $\sn$:
\begin{subequations}\label{eq:diffusion}
  \begin{equation}\label{eq:dtn}
    \partial_t\sn=-\bm{\nabla}\cdot\bm{\sJ}\,,
  \end{equation}
  where the constitutive relation is
  \begin{equation}\label{eq:constit}
    \bm{\sJ}=-\sth{\lambda}\bm{\nabla}\sth{\alpha}
    +\sqrt{\sth\lambda}\,\bm{\noise},
  \end{equation}
\end{subequations}
with the stochastic noise $\bm{\noise}$ given by
\begin{equation}
  \av{\noise^i(t_1,\bm x)\,\noise^j(t_2,\bm y)}
  =2\delta^{ij}\delta(t_1-t_2)\delta^{(3)}(\bm x-\bm y).
\end{equation}
$\sth\lambda=\lambda(\sn)$ is conductivity and $\sth\alpha=\alpha(\sn)$ is chemical
potential (in units of temperature).

To translate the generic multivariable stochastic system into a
stochastic hydrodynamic diffusion problem, we use the following dictionary:
\begin{subequations}\label{eq:dict}
  \begin{align}
    \delta_{ij} &\to \delta^{(3)}(\bm x - \bm y)
                  \equiv \delta_{\bm x\bm y},                  
    ~\q_i \to n(\bm x)\equiv n_{\bm x},\\
    S &\to \int_{\bm x} \left(s(n_{\bm x}) + \bar\alpha n_{\bm x}\right),\\
    S_{,i} &\to {\delta S}/{\delta n_{\bm x}} = - (\alpha(n_{\bm x}) - \bar\alpha),\\
    \M_{ij} &\to -\bm\nabla_{\bm x} \lambda(n_{\bm x}) \bm\nabla_{\bm x}
              \delta_{\bm{xy}},
  \end{align}
\end{subequations}
where $\alpha=-\partial s/\partial n$ and $\bar\alpha$ is a constant,
understood as the chemical potential (in units of
temperature) of the particle reservoir controlling the average
particle density  (similar to a Lagrange multiplier). Since
the indices $i$, $j$, etc.\ become continuous coordinates
$\bm x$, $\bm y$, etc., matrices $M$ and $H$ become
operators, which in Eq.~(\ref{eq:dict}) are expressed in terms of
their integral kernels. They are local, i.e., could be also written as
differential operators, i.e., $H=\bm\nabla\sqrt\lambda$ and
$M=-\bm\nabla\lambda\bm\nabla$.  Note that $\M=HH^T$, i.e., $\Omega=0$
for the diffusion problem. 

As in Eq.~(\ref{eq:qdot}), because the noise is multiplied
by a function of the stochastic variable $\sn$, the stochastic
equation [Eq.~\eqref{eq:diffusion}] is not well-defined as written. As
before, we can define the problem by specifying the equilibrium
distribution or, equivalently, the thermodynamic entropy $S$, as well as
the Onsager matrix/operator~$M$,  given in
Eq.~(\ref{eq:dict})\footnote{In higher
  orders of gradient expansion more parameters are needed~\cite{Jain:2020fsm}.}.

The role of expansion parameter allowing us to organize and
systematically truncate the infinite system of coupled equations for
correlation functions is played by a certain ratio of scales.
Hydrodynamics describes long wavelength modes of the field
$n_{\bm x}$, characterized by wave numbers $q\ll\xi^{-1}$, where $\xi$
is the microscopic correlation length. The noise is local, i.e.,
correlated on the scale $\xi$, and thus its effect on the
long wavelength modes involves averaging over a large number
$\mathcal O(q\xi)^{-3}$ of uncorrelated cells, suppressing
fluctuations (see, e.g., Refs.\cite{An:2019rhf,An:2019fdc}).  The
small parameter $(q\xi)^{3}$ plays a role similar to
parameter~$\varepsilon$ in Eq.~(\ref{eq:Gcneps}).  The smallness of~$H$,
suppressing the contribution of noise non-Gaussianity, is due to the
gradient in conservation equation [Eq.~\eqref{eq:dtn}] and is also
controlled by~$q\xi$.
Near the critical point, $\xi$ becomes large,
but as long as $\xi$ is much smaller than the system size,
hydrodynamic fluctuation theory applies.

Applying the dictionary in
Eq.~(\ref{eq:dict}), we can derive
evolution equations for the $n$-point correlators
\begin{equation}
  \label{eq:Gn}
  G_n(\bm x_1,\ldots,\bm x_n) \equiv \av{\dn(\bm x_1)\ldots\dn(\bm x_n)}.
\end{equation}
The equations for $G_n$ can then be
converted into corresponding equations for multipoint Wigner functions after
we introduce and define these objects.

\mbox{}

\section{Multipoint Wigner transform}

In hydrodynamics we consider fluctuations on a smoothly
varying background (see, e.g.,
Refs.~\cite{Akamatsu:2017,An:2019rhf,An:2019fdc}. As a result,
two separate scales characterize  fluctuation
correlators $G_n$. A shorter scale, corresponding to
wave number~$q$, characterizes the dependence on the separation
between points, while the dependence on the midpoint position occurs at a
much longer scale. 

The well-known method to take advantage of such a scale separation in
a two-point function is
to work with the Wigner transform (as in the derivation of kinetic
theory). In order to do this for $n$-point functions, we
need to generalize the Wigner transform, thus far only known for two-point
functions.

We propose to define the symmetric generalization of the Wigner
transform and its inverse as follows:
\begin{widetext}
  \begin{subequations}\label{eq:Wigner}
    \begin{equation}
      \label{eq:Wn}
      W_n(\bm{x},\bm{q}_1,\ldots,\bm{q}_n)
      \equiv
      \int d^3\bm{y}_1\ldots\int  d^3\bm{y}_n\,G_n\left(\bm{x}+{\bm{y}_1},\ldots,\bm{x}+{\bm{y}_n}\right)\delta^{(3)}\left(\frac{\bm{y}_1+\ldots+\bm{y}_n}{n}\right)
      e^{-i(\bm{q}_1\cdot\bm{y}_1+\ldots+\bm{q}_n\cdot\bm{y}_n)};
    \end{equation}
    \begin{equation}
      \label{eq:Wn-inv}
      G_n\left({\bm{x}_1},\ldots,{\bm{x}_n}\right) 
      =\int \frac{d^3\bm{q}_1}{(2\pi)^3}\ldots
      \int  \frac{d^3\bm{q}_n}{(2\pi)^3}\,W_n(\bm{x},\bm{q}_1,\ldots,\bm{q}_n)(2\pi)^3\delta^{(3)}\left({\bm{q}_1+\ldots+\bm{q}_n}\right)
      e^{i(\bm{q}_1\cdot\bm{x}_1+\ldots+\bm{q}_n\cdot\bm{x}_n)}\,.
    \end{equation}
  \end{subequations}
\end{widetext}
Note that, because of the $\delta$ function in Eq.~(\ref{eq:Wn}), the Wigner
function $W_n$ is invariant under the shift of all momenta $\bm{q}_i$
by the same vector. Effectively, there are only $n-1$ nonredundant
wave vector arguments in $W_n$ (i.e., the total number of nonredundant
arguments is the same as for $G_n$ itself).  Therefore,
it is not surprising that, in Eq.~(\ref{eq:Wn-inv}), to obtain $G_n$
we only need to evaluate $W_n$ for a set of $\bm q_i$'s that sum to
zero.  The same is true for all expressions that follow.

As an example, consider the case $n=2$. Then $\bm q_1=-\bm q_2\equiv \bm q$ and
\begin{multline}
  \label{eq:W2}
  W_2(\bm{x},\bm{q}_1,\bm{q}_2)= W_2(\bm{x},\bm q, -\bm q) \equiv
  W_2(\bm{x},\bm{q})
  \\
  \equiv \int d^3\bm{y}\,
  G_2(\bm{x}+\bm{y}/2,\bm{x}-\bm{y}/2) e^{-i\bm{q}\cdot\bm{y}}.
\end{multline}
This is the usual Wigner transform.  Because one of the $\bm q_i$'s is
always redundant, we shall adopt a simplified notation for $W_n$ by
dropping the redundant argument, as in Eq.~(\ref{eq:W2}). Since, in
this work, we consider Wigner functions symmetric w.r.t.\ their arguments, it
does not matter which argument is dropped.  Also note that the Wigner
transform of a generalized $n$-point $\delta$ function is unity for
all~$n$.

The Wigner transform of the partial derivative of~$G_n$ is given by
\begin{equation}
  \label{eq:dGn}
  \bm \nabla_{i} G_n
  \xlongrightarrow{\text{W.T.}}\left(i\bm q_{i} + \frac1n \bm\nabla_{\bm x}\right) W_n\,.
\end{equation}
In hydrodynamics, the gradient term is subleading to the term
proportional to $\bm q_i$. As a result, partial derivatives in
the equations for $G_n$ turn into factors of~$\bm q$ and partial differential
equations become ordinary differential equations. To simplify
notations, below we shall omit the spatial~$\bm x$ and the
time $t$ arguments for the
Wigner functions.

Applying the generalized Wigner transform  to the
evolution equations for $G_n$ we arrive at the
following evolution equations for $n=2,3,4$:
\begin{subequations}\label{eq:W_234}
     \begin{widetext}
  \begin{equation}\label{eq:W_234-a}
    \partial_tW_2(\bm{q}_1)
    =-\left[\gamma\bm{q}_1^2W_2(\bm{q}_2)+\lambda\bm{q}_1\cdot\bm{q}_2
       \right]_\qpermii
       \,,
     \end{equation}
     \begin{equation}\label{eq:W_234-b}
       \partial_tW_3(\bm{q}_1,\bm{q}_2)
       =-\left[\frac{1}{2}\gamma\bm{q}_1^2W_3(\bm{q}_2,\bm{q}_3)
         +\frac{1}{2}\gamma'\bm{q}_1^2W_2(\bm{q}_2)W_2(\bm{q}_3)
         +\lambda'\bm{q}_1\cdot\bm{q}_2W_2(\bm{q}_3)\right]_\qpermiii
       \,,
     \end{equation}
     \begin{multline}\label{eq:W_234-c}
       \partial_tW^\tc_{4}(\bm{q}_1,\bm{q}_2,\bm{q}_3)
       =-\left[
         \frac{1}{6}\gamma\bm{q}_1^2W^\tc_{4}(\bm{q}_2,\bm{q}_3,\bm{q}_4)
         +\frac{1}{2}\gamma'\bm{q}_1^2W_2(\bm{q}_2)W_3(\bm{q}_3,\bm{q}_4)
       \right.
    \\
    \left.
      +\frac{1}{6}\gamma''\bm{q}_1^2W_2(\bm{q}_2)W_2(\bm{q}_3)W_2(\bm{q}_4)
      +\frac{1}{2}\lambda'\bm{q}_1\cdot\bm{q}_2W_3(\bm{q}_3,\bm{q}_4)
      +\frac{1}{2}\lambda''\bm{q}_1\cdot\bm{q}_2W_2(\bm{q}_3)W_2(\bm{q}_4)
    \right]_\qpermiv
       ,
     \end{multline}
   \end{widetext}
 \end{subequations}
 where $\gamma=\lambda\alpha'$.
 Eq.~(\ref{eq:W_234}) is the main result of this work.

 It is
easy to map Eq.~(\ref{eq:W_234}) to diagrams in
Figs.~\ref{fig:diagram_notations} and~\ref{fig:diagram_equations} by
using the dictionary in Eq.~(\ref{eq:dict}) 
and noting that
$S_{,ij}\to-\alpha'$, $S_{,ijk}\to-\alpha''$,
$M_{ij}\to-\lambda \bm q_1\cdot\bm q_2$,
$M_{ij,k}\to -\lambda'\bm q_1\cdot\bm q_2$, etc.
Eq.~(\ref{eq:W_234}) describes the relaxation of the Wigner functions to
their equilibrium values which, as one can check, agree with
thermodynamics. Consistent with the underlying conservation laws,
the longer wavelength modes relax more slowly \footnote{See
  \hyperref[sec:suppl-mater-evol]{Supplemental Material} for an illustration of the evolution of Wigner functions governed by Eq.~(\ref{eq:W_234}).} at a rate proportional to
the square of their wave number.

Eq.~\eqref{eq:W_234} bears a certain similarity to equations in
Ref.~\cite{Mukherjee:2015swa} for cumulants in a uniform finite size
system (with $1/q$ in Eq.~(\ref{eq:W_234}) and the system size in
Ref.~\cite{Mukherjee:2015swa} playing similar roles). However, the
terms with derivatives of $\lambda$, related to multiplicative noise,
are absent in Ref.~\cite{Mukherjee:2015swa}, where $\lambda$ is a
constant. These terms are not negligible in general and, in
particular, near the liquid-gas-type
critical point where $\lambda$ is divergent~\cite{Hohenberg:1977}.


We considered the problem of the diffusion of a conserved
quantity and left the generalization to full stochastic
hydrodynamics, including pressure and flow, to future work. However, we
can anticipate some features of this full system based on what
we have already learned from the diffusion problem.

If we focus on the
fluctuations of the slowest hydrodynamic mode, that is, entropy per charge,
$m\equiv s/n$, at constant pressure, we should find a similar diffusion
equation with the substitution
\begin{equation}
  \label{eq:gamma,alpha}
  n\to  m,\quad
  \gamma \to \frac{\kappa}{c_p},\quad
  \alpha' \to \frac{n^2}{c_p}\,,
\end{equation}
where $\kappa$ is thermal conductivity and $c_p$ is heat capacity at
constant pressure.

The equation for a {\em two}-point function $\av{\delta m\delta m}$ was
derived by two different methods in
Refs.~\cite{Stephanov:2018hydro+,An:2019fdc}.  As expected, in the
regime we consider, $q\xi\ll1$, it coincides with our
Eq.~(\ref{eq:W_234-a}) upon substitution
Eq.~(\ref{eq:gamma,alpha}). 
One can argue that this should hold true for higher-point correlators
as well.  It would be interesting and important for applications to
heavy-ion collisions to combine the approach to non-Gaussianity
presented here with the relativistic treatment of fluctuations in
Ref.~\cite{An:2019fdc}. One can anticipate that, in addition to
Eq.~(\ref{eq:gamma,alpha}), the time derivatives will be replaced by
corresponding Liouville operators. Additional correlators will appear
but,  near the critical
point, the slowest and thus most out-of-equilibrium fluctuations will
be described by cumulants of $m$.  The validity of this argument
should be checked by direct derivation, which we defer to future work.

\section{Conclusions}
\label{sec:conclusions}

We found a systematically controllable hierarchy of equations
describing the evolution of higher-order, non-Gaussian cumulants of
fluctuations in a general multivariable stochastic system and
introduced a convenient diagrammatic representation.

We used this approach to tackle the problem of stochastic nonlinear
diffusion with  density-dependent conductivity, which involves
multiplicative noise. We introduced a generalization of a Wigner
transform to multipoint correlation function [Eq.~(\ref{eq:Wigner})] that allows us to take advantage of the separation of scales in
hydrodynamics and obtain evolution equations [Eq.~\eqref{eq:W_234}]. The
equations for the full system of hydrodynamic variables can be derived
along the same lines, and we defer this to future work.

It would also be interesting to extend the evolution
equations [Eqs.~\eqref{eqs:dt_Wc234} and \eqref{eq:W_234}] beyond the leading
order to loop diagrams and study the effects of ultraviolet
renormalization and long-time tails, as in, e.g.,
Ref.~\cite{An:2019fdc}, now involving multipoint correlations and
multiplicative noise.

The formalism and the results we present are very general and would
pertain to problems where nonlinearity and non-Gaussian fluctuations
are of interest --- from cosmology and astrophysics
\cite{KNOBLOCH198039,RevModPhys.15.1,1993ApJ...402..387L} to the
physics of electronic devices
\cite{PhysRev.108.541,PhysRevLett.90.206801,PhysRevE.84.011122},
glassy, granular, or colloidal materials \cite{PhysRevLett.81.3848},
and even finance \cite{PhysRevLett.89.098701}, to name only a few
examples.  Among the most immediate and practical applications is the
description of the evolution of the non-Gaussian measures of
fluctuations in heavy-ion collisions, which is crucial for the ongoing QCD
critical point search.

\acknowledgments

We thank M.~Hongo, D.~Pilipovic and Y.~Yin for helpful discussions and
comments.  This work is supported by the U.S. Department of Energy,
Office of Science, Office of Nuclear Physics, within the framework of
the Beam Energy Scan Theory (BEST) Topical Collaboration and grant
No. DE-FG0201ER41195.


\bibliographystyle{utphys}
\bibliography{references,references-2}



\newcommand\eqW{(\ref{eq:W_234})}
\newcommand\refMem{\cite{Berdnikov:1999ph,Mukherjee:2015swa}}


\newpage
\onecolumngrid 

\section{Supplemental material
}
\label{sec:suppl-mater-evol}

\twocolumngrid 

\subsection{Numerical solution}

To illustrate the main features of the evolution of Wigner
functions we solve Eq.~\eqW\ numerically in a simplified setup relevant
to critical point search in heavy-ion collisions. We consider the
system that expands along a trajectory which passes through the crossover
region near a critical point. The main features of the equation of state are due
to the singular behavior of the correlation length which we model
approximately by an ansatz which satisfies well-known scaling
properties, i.e., $\xi = r^{-\nu}f(Mr^{-\beta})$. A simple model of
the scaling function, $f=(1+x^2)^{-\nu/2\beta}$, will be more than
sufficient for our purposes.  The Ising variables $r$ (reduced
temperature) and $M$ (magnetization) are given by linear functions of
$T-T_c$ and $n-n_c$ respectively. We model expansion by assuming that $n$ is a
linear function of time~$t$, choosing $n(t=0)=n_c$. We also use the scaling
form of the inverse susceptibility $\alpha'\sim\xi^{-\gamma/\nu}$ and
conductivity $\lambda\sim\xi^{x_\lambda}$ with well-known critical
exponents. We use arbitrary units, since our purpose is to illustrate
the major qualitative features.


To further simplify the calculation without losing the main features
of the results, we consider an isotropic system such that
$W_n$ depend only on scalar invariants. We note that the number of
such {\em independent} invariants of $n$ vectors $\bm q_i$ which obey a
constraint $\sum_i\bm q_i=0$ is $n(n-1)/2$. We can choose them to be
$\bm q_i\cdot\bm q_j$, $i < j$, since 
$\bm q_i^2 = -\sum_{j\neq i} \bm q_i\cdot\bm q_j$.

\begin{figure}[H]
\vskip 1em
  \includegraphics[scale=0.55]{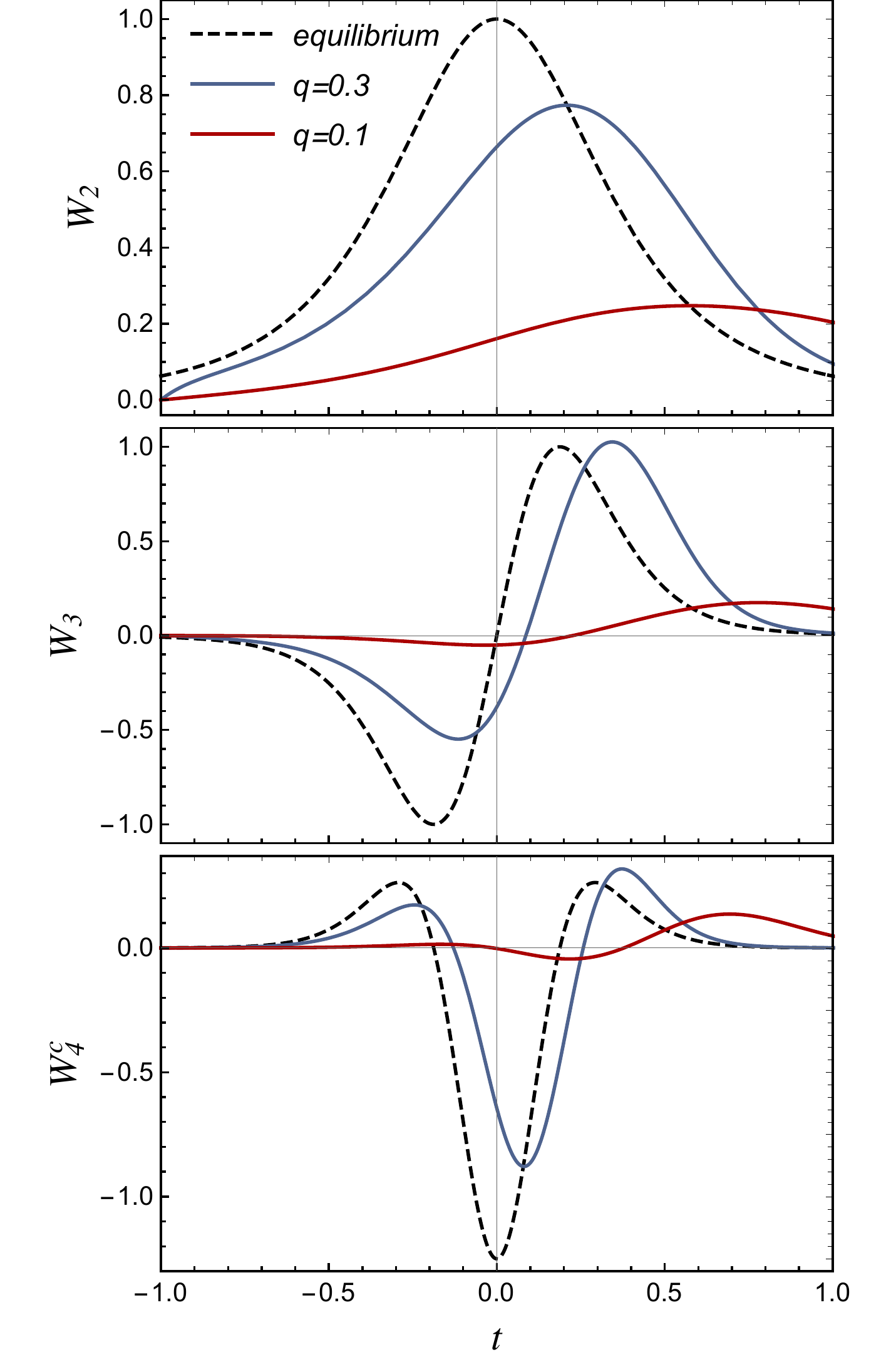}
\caption{The time dependence of Wigner functions in comparison with
  their equilibrium value (dashed line) as the system is driven by
  expansion through the critical region. The evolution strongly
  depends on the scale of the fluctuations given by $q$ (see text). }
\label{fig:ws}
\end{figure}

We shall demonstrate the dependence of the Wigner functions on time
and on the scale of $\bm q$. For that purpose we shall plot solutions
for a set of momenta for which all independent invariants
$\bm q_i\cdot\bm q_j$ are equal to the same value~$q^2$. 
The numerical solutions of Eq.~\eqW\ for two different representative
values of~$q$ are shown in Fig \ref{fig:ws}.

 One readily observes two
characteristic features. First, due to the relaxation nature of the
equations, the evolution of $W_n$ lags behind the change of their
equilibrium values (dashed curves) driven by the expansion of the
system through the critical region. Similar ``memory'' effects have been
observed in various models for cumulants in Refs.~\refMem. Second, unlike
previous approaches, Eq.~\eqW\ also describes the dependence of the memory
effects on the wave number $q$ of the fluctuations. As expected from the
conservation laws underlying hydrodynamics, the longer wavelength
modes relax slower, i.e., retain ``memory'' longer. This behavior has
important consequences for observation of critical fluctuations in
heavy-ion collisions. The strength of the observed signal depends on
the fluctuation scale probed. A more quantitative analysis
of such effects will require application of Eq.~\eqW\ in a more
realistic setting of a heavy-ion collision and we leave it to future
work.

\end{document}